\documentclass[prd,twocolumn,a4paper,superscriptaddress,floatfix]{revtex4}
\usepackage{graphicx,color}
\usepackage{amsmath}
\begin{document}

\newcommand{\be}{\begin{equation}}
\newcommand{\ee}{\end{equation}}
\newcommand{\bq}{\begin{eqnarray}}
\newcommand{\eq}{\end{eqnarray}}
\newcommand{\bsq}{\begin{subequations}}
\newcommand{\esq}{\end{subequations}}
\newcommand{\bc}{\begin{center}}
\newcommand{\ec}{\end{center}}

\title{Cosmic string evolution with a conserved charge}
\author{M. F. Oliveira}
\email[Electronic address: ]{up070308025@alunos.fc.up.pt}
\affiliation{Centro de Astrof\'{\i}sica, Universidade do Porto, Rua das Estrelas, 4150-762 Porto, Portugal}
\affiliation{Faculdade de Ci\^encias, Universidade do Porto, Rua do Campo Alegre 687, 4169-007 Porto, Portugal}
\author{A. Avgoustidis}
\email[Electronic address: ]{A.Avgoustidis@damtp.cam.ac.uk}
\affiliation{School of Physics and Astronomy, University of Nottingham, University Park, Nottingham NG7 2RD, UK}
\author{C. J. A. P. Martins}
\email[Electronic address: ]{Carlos.Martins@astro.up.pt}
\affiliation{Centro de Astrof\'{\i}sica, Universidade do Porto, Rua das Estrelas, 4150-762 Porto, Portugal}

\date{24 January 2012}

\begin{abstract}
Cosmic strings with degrees of freedom beyond the standard Abrikosov-Nielsen-Olesen or Nambu-Goto  strings are ubiquitous in field theory as well as in models with extra dimensions, such as string theoretic brane inflation scenarios. Here we carry out an analytic study of a simplified version of one such cosmic string model. Specifically, we extend the velocity-dependent one-scale (VOS) string evolution model to the case where there is a conserved microscopic charge on the string worldsheet. We find that whether the standard scale-invariant evolution of the network is preserved or destroyed due to the presence of the charge will crucially depend on the amount of damping and energy losses experienced by the network. This suggests, among other things, that results derived in Minkowski space (field theory) simulations may not extend  to the case of an expanding universe.
\end{abstract}

\maketitle


\section{Motivation}

Cosmic strings \cite{VSH,HindKib}, line-like topological defects 
that can be produced in cosmological phase transitions, provide a 
valuable tool for constraining models of the early universe. Having 
a wide range of potential observational signals \cite{CopPogVach}, which 
depend directly on their microphysical properties, they can be used to 
constrain high-energy physics parameters from cosmological observations.
In particular, the magnitude of their observational signatures is mainly 
determined by the energy scale of the corresponding symmetry breaking 
transition. Interest in the study of cosmic strings has been renewed in 
recent years \cite{KibbleRev} with the realisation that string-like defects 
are generic in a wide range of fundamental models, including 
Supersymmetric Grand Unified Theories \cite{JeRoSak} and Brane 
Inflation models in String Theory \cite{BMNQRZ,SarTye,DvalVil,CopMyePolch}.  
In the latter case, there is even the exciting possibility of constraining (within 
a given model) fundamental parameters like the string coupling, string scale, 
and warping/compactification scales from cosmological 
observations \cite{PACPS,ACMPPS,CopPogVach}.     

In order to be able to make this quantitative link between late time
observations and early universe physics, one must therefore 
be able to model the evolution of string networks from the 
time of their formation until they are observed.  Work in this direction 
has mostly focused on the simplest, structureless strings of the 
Abrikosov-Nielsen-Olesen type and their corresponding Nambu-Goto 
thin string approximations. In recent years, more complex models
have been developed (both on the numerical \cite{CopSaf,UrrVil,HindSaf,SakSto} 
and analytical \cite{TWW,AvgShell,AvgCop} fronts), allowing for the 
formation of zipping interactions between strings of different tensions, 
in order to describe the richer nature of the so called FD-networks 
\cite{PolchIntro} that appear in Brane Inflation models.  

However, even networks of a single type of string can have non-trivial 
internal structure, generally carrying additional degrees of freedom on 
the string worldsheet. This is the generic situation in models with extra 
dimensions, where there is a proliferation of scalar fields that can couple 
to (and condense on) the strings.  Cosmic strings with additional worldsheet 
degrees of freedom (scalar charges, currents, fermionic zero-modes) 
have been extensively studied in field theory and supergravity, see for 
example \cite{Witten,CalJef,Vorton1,GanLaz,CarPet,BvdBDD}.  
One important result that emerged from these studies was the realisation that 
the presence of currents can lead to the formation of stable cosmic string loops, 
known as vortons \cite{Vorton1,Vorton2}, which could therefore dominate the energy 
density of the universe.  This has allowed to obtain strong bounds on models 
producing vortons \cite{BCDT,SUPERC2,Vorton3}. 

To date, the cosmological evolution of such string networks remains largely 
unexplored. It is therefore desirable to understand how these additional degrees 
of freedom can be described macroscopically, and how their presence affects the 
behaviour and cosmological consequences of the corresponding string networks. 
In a recent paper \cite{NAMU}, the evolution of semilocal strings \cite{VachAch} was 
described analytically and it was found that the spectrum of scaling solutions has 
a much richer structure than in ordinary cosmic strings.  The presence of worldsheet 
degrees of freedom would also modify significantly the structure of scaling network 
solutions and can be expected to obstruct scaling, possibly leading to string frustration.        

Here, we focus on the case of cosmic strings with a conserved charge living on the string worldsheet.  We
extend the VOS model \cite{MS1,MS2,MS3,MS4} to describe this case analytically, 
and study the effect of the charge on the evolution of the string network. 
A recent related example, for the case of domain walls, comes from the work of Battye \textit{et al.} \cite{BATTYE1} (but see also \cite{BATTYE2} for some caveats), who suggest that with a conserved, homogeneous and localised Noether charge the number of domain walls does not scale in the usual way, in which case they could provide some contribution to the dark energy suggested by cosmological observations. These results are based on relatively small numerical simulations in $2+1$ dimensions and without cosmological expansion, so extrapolating them to a cosmological context requires some care. Nevertheless, they highlight the need for a better macroscopic description of these processes. As we shall see, our analysis will provide some possible explanations for these results.


\section{Microscopic model}

Here we review the basics of the microscopic model that we will be using. This model is described in detail in Ref.~\cite{SUPERC2}. A comprehensive description of the general formalism can be found in \cite{CARTERB}.

\subsection{Generic chiral models}\label{GenChir}

For strings carrying conserved currents, we consider the Witten-Carter-Peter (chiral) model, which implicitly makes use of the fact that in two dimensions a conserved current can be written as the derivative of a scalar field. The dynamics is described by the Witten action
\begin{widetext}
\begin{equation}\label{SWitten}
S_W=\int\sqrt{-\gamma}\left[-\mu_0+\frac{1}{2}\gamma^{ab}\phi_{,a}\phi_{,b} -qA_\mu x^\mu_{,a}
\frac{\tilde\epsilon^{ab}}{\sqrt{-\gamma}}\phi_{,b}\right]d^2\sigma -\frac{1}{16\pi}\int
d^4x\sqrt{-g}F_{\mu\nu}F^{\mu\nu}\,,
\end{equation}
\end{widetext}
where the four terms are respectively the usual Nambu-Goto term ($\mu_0$ being 
the string tension and $\gamma$ the pullback of the background metric $g$ on the worldsheet), 
the inertia of the charge carriers described by the scalar $\phi$, the current coupling to the 
electromagnetic potential $A_\mu$, and the kinetic term for the electromagnetic field.   
Worldsheet indices are denoted by $a,b\in \{0,1\}$ and we will take $\sigma^0$ to be 
the timelike coordinate, while $\sigma^1\equiv\sigma$ will be spacelike; $\tilde\epsilon^{ab}$ 
is the alternating tensor in two dimensions. Note that this action applies to both the bosonic 
and the fermionic case.

We are essentially interested in the chiral limit of this model, that is 
(taking dot/prime to denote differentiation with respect to the 
timelike/spacelike worldsheet coordinate $\sigma^0$/$\sigma$):
\begin{equation}\label{phiprphidot}
\phi'{}^2=\epsilon^2 {\dot\phi}^2\,,
\end{equation}
where $\epsilon$ is the scalar
\be
\epsilon\equiv \frac{-{x'}^2}{\sqrt{-\gamma}}\,,
\ee
giving the string energy per unit coordinate length.
Let us now consider an FRW background
\be
ds^2=a^2(d\tau^2 - d{\bf x}^2) 
\ee
and choose the standard gauge 
\begin{equation}
\sigma^0=\tau\, , \qquad {\dot{\bf x}}\cdot{\bf x}'=0 \,,
\end{equation} 
in which the scalar $\epsilon$ becomes:
\be
\epsilon={\left(\frac{{\bf
  x}^{\prime}{}^2}{1-{\dot{\bf x}}^2}\right)}^{1/2}\,.
\ee
Then, in the chiral limit, introducing the simplifying function 
$\Phi$ defined as
\begin{equation}\label{Phi}
\Phi(\phi)=\frac{{\dot\phi}^2}{\mu_0 a^2(1-{\bf {\dot x}}^2)}\, ,
\end{equation}
the microscopic equations of motion take the form~\cite{SUPERC2}:
\begin{equation}\label{timecpt}
\left[\epsilon\left(1+\Phi\right)\right]{\dot{}} +\frac{\epsilon}{\ell_d}{\dot {\bf x}}^2 =\Phi'-2\frac{{\dot a}}{a}\epsilon\Phi \,,
\end{equation}
and
\begin{widetext}
\begin{equation}\label{spacecpt}
\epsilon\left(1+\Phi\right) {\ddot {\bf x }}+\frac{\epsilon}{\ell_d}(1-{\dot {\bf x}}^2) {\dot {\bf x}}=\left[\left(1-\Phi\right) \frac{{\bf x}'}{\epsilon}\right]'+\left({\dot \Phi}+2\frac{{\dot a}}{a}\Phi\right){\bf x}'+2 \Phi{\dot{\bf x}}' \,,
\end{equation}
\end{widetext}
where for simplicity we have introduced the damping length
\begin{equation}
\frac{1}{\ell_d}=a\left(2H+\frac{1}{\ell_{\rm f}}\right)\, ,
\end{equation}
describing Hubble friction ($H=a^{-1}(da/dt)={\dot a}/a^2$), but also allowing 
for an additional friction mechanism with characteristic lengthscale 
$\ell_{\rm f}$.

Two generic useful relations are
\begin{equation}\label{epsilondot}
\frac{\dot\epsilon}{\epsilon}=\frac{{\dot {\bf x }}\cdot{\ddot {\bf x }}}{1-{\dot {\bf x}}^2}-\frac{{\dot {\bf x }}\cdot{\bf x}''}{{\bf x'}{}^2}
\end{equation}
and
\begin{equation}\label{epsilonprime}
\frac{\epsilon'}{\epsilon}=\frac{{\bf x}'\cdot{\bf x}''}{{\bf x'}{}^2}-\frac{{\bf x}'\cdot{\ddot {\bf x }}}{1-{\dot {\bf x}}^2}\,.
\end{equation}
The curvature vector $d^2{\bf x }/ds^2$, with $ds=\sqrt{{\bf x}^\prime{}^2}d\sigma$, satisfies:
\begin{equation}\label{curvature}
\frac{d^2{\bf x }}{ds^2}=\frac{{\bf x}''}{{\bf x'}{}^2}-\frac{({\bf x'}\cdot{\bf x}''){\bf x'}}{{\bf x'}{}^4}\,.
\end{equation}
Thus, a curvature radius, $R$, can be defined locally via
\begin{equation}
\frac{\dot {\bf x }}{\epsilon(1-{\dot {\bf x}}^2)}\cdot\left(\frac{{\bf x'}}{\epsilon}\right)'=-\frac{{\bf x'}\cdot{\dot {\bf x' }}}{{\bf x'}{}^2} = \frac{{\bf x}''\cdot{\dot {\bf x }}}{{\bf x'}{}^2}=\frac{a}{R}({\dot {\bf x }}\cdot{\bf u})\,,
\end{equation}
where we have introduced a unit vector $\bf u$ in the direction of the curvature vector.

The worldsheet charge and current densities are respectively given by
\begin{equation}
\rho_{w}=q\epsilon{\dot \phi}\, ,
\end{equation}
and
\begin{equation}
j_{w}=q\frac{\phi'}{\epsilon}\, ,
\end{equation}
while the total energy of a piece of string is given by
\begin{equation}\label{TotEnergy}
E=\mu_0 a\int\left(1+\Phi\right)\epsilon d\sigma=E_s+E_\Phi\, .
\end{equation}
We can immediately interpret the energy (\ref{TotEnergy}) as being 
split in an obvious way into a string component and a charge component. 
Defining a macroscopic charge as the average of $\Phi$ over the
string worldsheet, we then have:
\begin{equation}\label{macrocharge}
Q=\langle\Phi\rangle\equiv \frac{\int\Phi\epsilon d\sigma}{\int\epsilon d\sigma}=\frac{E_\Phi}{E_s}\,.
\end{equation}
This interpretation will be relevant below.

Introducing the network string density $\rho_s$ such 
that $E_s\propto \rho_s a^3$ and defining the correlation length 
$\xi$ by
\be\label{xi_def}
\rho_s=\frac{\mu_0}{\xi^2} \,,
\ee
the evolution equations have the form
\begin{equation}
\frac{\dot E_s}{E_s}=\frac{\dot\rho_s}{\rho_s}+3\frac{\dot a}{a}=-2\frac{\dot \xi}{\xi}+3\frac{\dot a}{a}= \frac{\dot a}{a}+
\left\langle\frac{\dot\epsilon}{\epsilon}\right\rangle\,,
\end{equation}
\begin{equation}
v^2=\langle{\dot {\bf x}}^2\rangle\,,\quad v{\dot v}=\langle{\dot {\bf x}}\cdot{\ddot {\bf x}}\rangle
\end{equation}
and
\begin{equation}
{\dot Q}=\langle{\dot\Phi}\rangle\,.
\end{equation}
Two further (not independent) useful relations are
\begin{equation}
\frac{\dot E_\Phi}{E_\Phi}=\frac{\dot Q}{Q}+\frac{\dot E_s}{E_s}\,,
\end{equation}
and
\begin{equation}
\frac{\dot E}{E}=\frac{\dot E_s}{E_s}+\frac{\dot Q}{1+Q}=\frac{\dot\rho}{\rho}+3\frac{\dot a}{a}=-2\frac{\dot L}{L}+3\frac{\dot a}{a}\,,
\end{equation}
where $\rho\propto Ea^{-3}$ is the energy density and the lengthscale $L$
is defined by: 
\be\label{L_def}
\rho=\frac{\mu_0}{L^2} \,.
\ee 
With these, one can in principle proceed to average this model. An additional difficulty 
which is absent in the case of Nambu-Goto strings is the appearance of a term proportional to 
(${\bf x}'\cdot{\bf x}''$).  This factor is not expected to be zero even though (${\bf x }'\cdot{\bf u}$) 
is, see equation (\ref{curvature}). This will be discussed in more detail below.

Finally, note that in the case of the VOS model for plain Nambu-Goto strings one assumes that the network has a single characteristic length scale, so that $R=L=\xi$. This is no longer true in the charged case, but we will still assume that $R=\xi$, while $L$ is now only a measure of the total energy in the network.

\subsection{Conserved microscopic charge case}

In our particular case we have
\begin{equation}
\rho_w=q\epsilon{\dot\phi}=q\phi'=const.
\end{equation}
and therefore
\begin{equation}
\Phi=\frac{\varphi_0^2}{a^2 {\bf x'}{}^2}\,,
\end{equation}
with $\varphi_0$ being a constant. 
Now, by simple differentiation one finds that $\Phi$ evolves as follows
\begin{equation}\label{phidot}
{\dot\Phi}+2\frac{\dot a}{a}\Phi=2\Phi\frac{{\dot {\bf x }}\cdot{\bf x}''}{{\bf x'}{}^2}
\end{equation}
and
\begin{equation}\label{phiprime}
\Phi'+2\Phi\frac{{\bf x}'\cdot{\bf x}''}{{\bf x'}{}^2}=0\,.
\end{equation}

The conserved charge assumption simplifies some of the above equations. If one expands Eq. (\ref{epsilondot}) by successively substituting in Eqs. (\ref{timecpt}-\ref{spacecpt}) and then Eqs. (\ref{phidot}-\ref{phiprime}) one obtains
\begin{equation}\label{chargecond1}
{\ddot {\bf x}}\cdot{\bf x}'=\frac{{\bf x}'\cdot{\bf x}''}{\epsilon^2}\,,
\end{equation}
and inserting this in Eq. (\ref{epsilonprime}) yields
\begin{equation}\label{chargecond2}
\epsilon'=0\,.
\end{equation}
An alternative way to see this is to substitute Eq. (\ref{spacecpt}) into Eq. (\ref{epsilonprime}) and then use Eqs. (\ref{phidot}-\ref{phiprime}). This leads to
\begin{equation}
\frac{2\Phi}{1+\Phi}\frac{\epsilon'}{\epsilon}=0\,,
\end{equation}
thus if $\Phi\neq0$ we must have $\epsilon'=0$.


\section{Interlude: Nambu-Goto Strings in Extra Dimensions}

The above formalism describing strings with currents is also analogous to 
the effective description of structureless, Nambu-Goto strings evolving in 
the presence of compact extra dimensions.  Indeed, integrating out the extra 
dimensions gives rise to worldsheet scalars which geometrically correspond 
to string positions in the internal manifold and behave like currents in the 
low energy worldsheet theory.  

The evolution and dynamics of strings moving in spacetimes with extra compact
dimensions have been studied in Ref.~\cite{XDIMR}.  Consider the 
following `augmented' FRW metric
\be\label{EDmetric}
  ds^2=dt^2 - a(t)^2 d{\bf x}^2 - b(t)^2 d{\bf l}^2 \,,
\ee
where the internal manifold (coordinates $\bf l$) has been taken for simplicity 
to be toroidal with scalefactor $b(t)$. Now let the vector $\bf l(\tau,\sigma)$ denote 
the string position in this internal manifold.  In the simplest static case, $b(t)=1$, 
the string energy is:
\bq
  \varepsilon_e &\equiv& \frac{-x^{\prime}{}^2}{\sqrt{-\gamma}}={\left(\frac{{\bf
  x}^{\prime}{}^2+{\bf l}^{\prime}{}^2/a^2}{1-{\dot{\bf x}}^2-{\dot
  {\bf l}}^2/a^2}\right)}^{1/2} \label{EDepsilon} \\
  &\simeq& {\left(\frac{{\bf
  x}^{\prime}{}^2}{1-{\dot{\bf x}}^2}\right)}^{1/2}\left( 1 + 1/2 
  \frac{{\bf l}^{\prime}{}^2}{a^2{\bf x}^{\prime}{}^2} + 
  1/2 \frac{{\dot{\bf l}}^2}{a^2(1-{\dot{\bf x}}^2)} \right) \,, 
   \nonumber
 \eq
where we have expanded to linear order in 
${\bf l}^{\prime}{}^2/(a^2 {\bf x}^{\prime}{}^2)$ 
and ${\dot{\bf l}}^2/[a^2(1-{\dot{\bf x}}^2)]$. 

The analogy to the above discussion is apparent: 
the first factor in the last line of (\ref{EDepsilon}) is 
simply $\epsilon$, while the terms in the second factor
are of the form of equations (\ref{Phi}) and (\ref{phiprphidot}).   
Indeed, setting ${\dot{\bf l}}^2/[a^2(1-{\dot{\bf x}}^2)]=\Phi$, 
and considering the analogue of the `chiral limit' 
$\phi^{\prime}{}^2=\epsilon^2 \dot\phi^2$ 
discussed above, we then have in this case
\be\label{chiral_l}
{\bf l}^{\prime}{}^2=\frac{ {\bf x}^{\prime}{}^2 }{ (1-{\dot{\bf x}}^2) } {\dot{\bf l}}^2 \,,
\ee
and equation (\ref{EDepsilon}) becomes
\[
\varepsilon_e \simeq \epsilon (1+\Phi)
\] 
in direct analogy to equation (\ref{TotEnergy}).
In addition, the equation of motion for $\varepsilon_e$
\be
\frac{\dot\varepsilon_e}{\varepsilon_e}=-\frac{\dot a}{a}\left[1+{\dot{\bf x}}^2
-{\bf x}^{\prime}{}^2/\varepsilon_e^2\right]
\ee  
with $\varepsilon_e \simeq \epsilon (1+\Phi)$ becomes:
\be
\left[\epsilon\left(1+\Phi\right)\right]{\dot{}} = -2\frac{{\dot a}}{a}\epsilon({\dot{\bf x}}^2+\Phi) 
+ \mathcal{O}(\Phi^2) \,,
\ee 
which is equation (\ref{timecpt}) for $\Phi^\prime=0$ to linear order in $\Phi$. 

Note that this analogy only becomes quantitative
in the limit where ${\bf l}^{\prime}{}^2/(a^2 {\bf x}^{\prime}{}^2)$ 
and ${\dot{\bf l}}^2/[a^2(1-{\dot{\bf x}}^2)]$ are small.  The embedding 
fields ${\bf l}\equiv l^m$ ($m=1,...,D$), describing the string positions in 
the $D$ compact extra dimensions, are worldsheet scalars which in general 
have richer dynamics than in action (\ref{SWitten}).  In particular, their 
kinetic structure is governed by the relativistic action Nambu-Goto action 
in $4+D$ spacetime dimensions:
\be
S=-\mu_0 \int \sqrt{-{\rm det} \left(g^{(4)}_{\mu\nu} \partial_a x^\mu 
  \partial_b x^\nu + g^{(D)}_{\ell m} \partial_a l^\ell \partial_b l^m\right) } \,,
\ee   
where $g^{(4)}_{\mu\nu}$ is the (4-dimensional) spacetime metric and 
$g^{(D)}_{\ell m}$ the metric in the $D$ compact dimensions.  Factoring 
out $\gamma^{(4)}_{ab}\equiv g^{(4)}_{\mu\nu} \partial_a x^\mu \partial_b x^\mu$ 
yields:
\be\label{Sfactorised}
S=-\mu_0 \int \sqrt{-{\rm det}\gamma^{(4)}_{ab}} 
  \sqrt{ {\rm det}(\delta^a_b+\gamma_{(4)}^{ac}  
  \, \partial_c l^\ell\partial_b l^m \, g^{(6)}_{\ell m})}  \,, 
\ee   
that is, the general action for the worldsheet scalars has non-trivial 
kinetic structure.  However, for small $\dot l$ and $l^\prime$, 
the second factor in (\ref{Sfactorised}) can be linearised, and 
with the additional `chiral' condition (\ref{chiral_l}) one recovers 
the first two terms of the Witten action (\ref{SWitten}), for $D$ 
scalars, $l^m$, subject to (\ref{chiral_l}).


\section{Loop solutions}

Simple exact loop solutions can be used to further our understanding of the role of the various terms on the evolution of the strings, and also determine the accuracy of averaged (macroscopic) quantities in relation to microscopic quantities. The simplest loop solution is a circular one, which has the form
\begin{equation}
\ {\bf x}={r(\tau) (\cos{\theta}, \sin{\theta}, 0)}
\end{equation}
with
\begin{equation}\label{epsPhiloop}
\epsilon=\frac{r}{\sqrt{1-\dot{r}^2}}\,,\quad \Phi=\frac{\varphi_0^2}{a^2r^2}\,,
\end{equation}
and
\begin{equation}
E=\mu_0 a\int\left(1+\frac{\varphi^2}{a^2r^2}\right)\epsilon d\sigma\,.
\end{equation}

For this ansatz Eqs.~(\ref{timecpt}) and (\ref{spacecpt}) are equivalent, and either of them can then be written
\begin{equation}\label{circularloop}
(1+\Phi) \ddot{r}+(1- \dot{r}^2)\left[2\frac{\dot a}{a}{\dot r}+(1-\Phi)\frac{1}{r}\right]=0\,,
\end{equation}
while the evolution equation for $\Phi$ has the form
\begin{equation}\label{circularloop_charge}
\dot{\Phi}=-2\Phi\left(H+\frac{\dot{r}}{r}\right)\,,
\end{equation}
which is in fact trivial given the definition of $\Phi$ in (\ref{epsPhiloop}).

We start by examining the case with no expansion ($H=0$). The evolution equations, 
Eqs. (\ref{timecpt}) and (\ref{spacecpt}), now simplify to
\begin{equation}
\epsilon (1+\Phi)=const. =1\,
\end{equation}
and
\begin{equation}
\ddot{r}+(1-\Phi^2) r = 0\,.
\end{equation}
Alternatively, they can be written as
\begin{equation}
\dot{r}^2+r^2=1-2\varphi_0^2-4\frac{\varphi_0^4}{r^2}\,
\end{equation}
and
\begin{equation}\label{chargeforce}
\ddot{r}+r = \frac{\varphi_0^4}{r^3}\,.
\end{equation}

These have periodic (oscillatory) solutions, but unlike the case of circular loops in the plain Nambu-Goto case, the presence of a charge ensures that these loops never collapse to zero size and that their microscopic velocity is always less than unity. Indeed, the maximum velocity is
\begin{equation}
\dot{r}^2_{\rm max}=1-4\varphi_0^2\,.
\end{equation}
The limiting behaviour corresponds to a static solution,
\begin{equation}
r=\varphi_0=\frac{1}{2}\,,\quad \dot{r}=0\,;
\end{equation}
in this solution the energy of the loop is equally divided, half of it being in the string itself and the other half in the current. In the general case, the average loop velocity (over one oscillation period) is
\begin{equation}
\langle\dot{r}^2\rangle=\frac{1}{2}(1-4\varphi_0^2)\,,
\end{equation}
and the average fraction of the loop's total energy in the current is
\begin{equation}
\left\langle\frac{E_{\Phi}}{E}\right\rangle=\varphi_0\,;
\end{equation}
naturally the rest of the energy corresponds to the (bare) string. Finally the macroscopic charge of these loops, given by Eq. (\ref{macrocharge}), is
\begin{equation}
Q=\frac{\varphi_0}{1-\varphi_0}\,;
\end{equation}
notice that this is unity in the static limit ($\varphi_0=1/2$).
The static solution we discussed above still exists in an expanding universe. In this case it corresponds to
\begin{equation}
ar=\varphi_0\,,\quad \dot{r}=0;
\end{equation}
such a loop will have a physical radius
\begin{equation}
R_{\rm phys}=\frac{ar}{\sqrt{1-\dot{r}^2}}=\varphi_0\,.
\end{equation}

\begin{figure}
\includegraphics[width=3.3in,keepaspectratio]{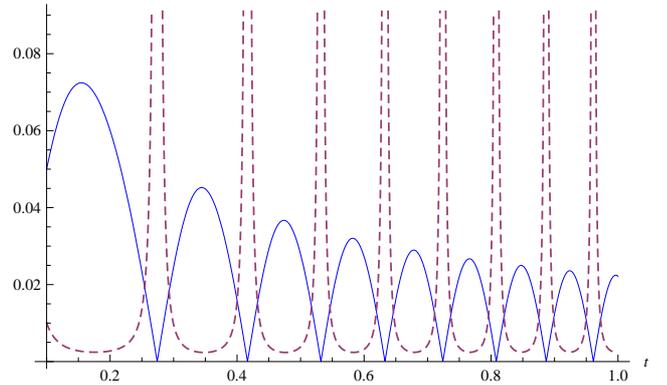}
\caption{\label{fig01} The time evolution of a particular loop solution in an expanding universe. The solid (blue) curve shows $r$, while the dashed (red) one shows the force $\varphi_0^4/r^3$ due to the charge, Eq. (\ref{chargeforce}). As the loop shrinks and $r$ approaches zero, the force provides a `boost' causing it to expand again.}
\end{figure}

Solving the differential equation for r and integrating over one period, we can obtain
\begin{equation}
\frac{\langle\dot{\bf{x}}^2\rangle^2}{\langle\dot{\bf{x}}^4\rangle}=\frac{2}{3}
\end{equation}
which is the same as in the case without charge. This result is important because the fact that it is always of order unity allows us to, every time a microscopic quantity such as $\dot{\bf x}^4$ appears in an expression inside an integral, average it to $v^4$ to a good approximation. In the general case with expansion, Eq. (\ref{circularloop}) cannot be solved analytically, but numerically we found that this result still holds to a good approximation for a wide range of initial conditions. Fig.~\ref{fig01} shows the behaviour of a particular solution, for which we find
\begin{equation}
\frac{\langle\dot{\bf{x}}^2\rangle^2}{\langle\dot{\bf{x}}^4\rangle}\sim0.6622\,.
\end{equation}

As can be seen in Fig.~\ref{fig01}, the charge in the loops keeps them from reaching 
$r=0$ by providing a `boost' when $r$ is approaching zero. It then causes $r$ to 
grow again (to a lower value than before), while it returns to its previous value.
Naturally, these simple loop solutions satisfy Eqs. (\ref{chargecond1}-\ref{chargecond2}), and indeed one can check that this is still the case for Kibble-Turok \cite{KIBTUR} and Burden \cite{BURDEN} loops.


\section{Macroscopic equations}

We can now proceed and look at the averaged evolution equations in our conserved microscopic charge case.

\subsection{General dynamical equations}

The total energy of the string is given by Eq. (\ref{TotEnergy})
and therefore we can define two characteristic lengths for the string, 
as in section \ref{GenChir}; the usual correlation length $\xi$ associated with the 
\emph{string} energy
\begin{equation}
E_s=\mu a \int{\epsilon d\sigma}\propto \rho_s a^3 \,,
\end{equation}
through Eqn. (\ref{xi_def}), and the lengthscale $L$ associated with the 
\emph{total} energy $E\propto \rho a^3$, see Eqs. (\ref{TotEnergy}), (\ref{L_def}).
Taking the time derivative of the previous equation, one easily finds the evolution equation for $\xi$
\begin{equation}\label{macro1}
2\frac{\dot{\xi}}{\xi}=2H\left(1+\frac{v^2}{1+Q}\right)+\frac{2Q}{1+Q} \frac{kv}{R}-\left\langle 
\frac{\Phi'}{\epsilon (1+\Phi)}\right\rangle\,.
\end{equation}
Here, $k$ is the so-called `momentum parameter' quantifying the average angle between 
the curvature vector and the velocity of string segments in the network.  It thus provides a measure 
of the small-scale-structure on strings and can range from 0 (wiggly strings in flat space) to order
unity (smooth strings) \cite{MS3}.

Correspondingly, the evolution equation for L is
\begin{equation}\label{macro2}
2\frac{\dot{L}}{L}=2H\left(1+\frac{v^2+Q}{1+Q}\right)-\left\langle \frac{\Phi'}{\epsilon 
(1+\Phi)}\right\rangle\,.
\end{equation}
As was defined above, Eqn. (\ref{macrocharge}), the macroscopic charge $Q$ 
is the averaged microscopic charge $\langle\Phi\rangle$, which is also the 
ratio between the string energy $E_s$ and the `charge' energy $E_\Phi$.  
Differentiating (\ref{macrocharge}), we find that its evolution equation is
\begin{equation}\label{macro3}
\frac{\dot{Q}}{Q}=2\left(\frac{kv}{R}-H\right)
\end{equation}
Finally, the string velocity is defined by
\begin{equation}
v^2=\langle{\dot{\bf{x}}^2}\rangle=\frac{\int\dot{\bf{x}}^2\epsilon d\sigma}{\int\epsilon d\sigma} \,,
\end{equation}
and taking time derivatives on both sides,  we arrive at the evolution equation 
\begin{equation}\label{macro4}
\dot{v}=\frac{1-v^2}{1+Q}\left[\frac{k}{R}(1-Q)-2Hv+\frac{1+Q}{v}\left\langle \frac{\Phi'}{\epsilon (1+\Phi)}
\right\rangle\right]
\end{equation}
However, equations (\ref{macro1}), (\ref{macro2}) and (\ref{macro3}) are not independent: there 
is a consistency relation between $\xi$ and $L$
\begin{equation}
E_s=\frac{E}{1+Q} \longrightarrow \xi=L \sqrt{1+Q}
\end{equation}
Therefore, the equations are related by
\begin{equation}\label{consisteq}
2\frac{\dot{L}}{L}=2\frac{\dot{\xi}}{\xi}-\frac{\dot{Q}}{1+Q}\,,
\end{equation}
which is verified, so the three equations are consistent.

\subsection{Ansatz for the charge gradient}

As previously mentioned, in order to proceed we now need to deal with the (${\bf x}'\cdot{\bf x}''$) term 
coming from $\Phi'$.  Referring to Eq. (\ref{phiprime}), dimensional analysis suggests an ansatz of the form
\begin{equation}\label{Phiprimeans}
\left\langle \frac{\Phi'}{\epsilon (1+\Phi)}\right\rangle=-s\frac{v}{R}\frac{2Q}{1+Q}
\end{equation}
where $s$ is (at least, to a first approximation) a constant.

Using (\ref{Phiprimeans}) and noting our earlier identification $R=\xi$, our evolution equations become:
\begin{equation}\label{macro10}
2\frac{\dot{\xi}}{\xi}=2H\left(1+\frac{v^2}{1+Q}\right)+\frac{2Q}{1+Q} \frac{(k+s)v}{\xi}\,
\end{equation}
\begin{equation}\label{macro20}
2\frac{\dot{L}}{L}=2H\left(1+\frac{v^2+Q}{1+Q}\right)+\frac{2Q}{(1+Q)^{3/2}} \frac{sv}{L}\,
\end{equation}
\begin{equation}\label{macro30}
\frac{\dot{Q}}{Q}=2\left(\frac{kv}{\xi}-H\right)\,
\end{equation}
\begin{equation}\label{macro40}
\dot{v}=\frac{1-v^2}{1+Q}\left[\frac{k}{\xi}\left(1-Q(1+2s/k)\right)-2Hv\right]\,.
\end{equation}

We will assume a critical-density universe with generic expansion rates of the form
\begin{equation}
a\propto t^\lambda\,,\quad H=\frac{\lambda}{t}\,,
\end{equation}
and look for scaling solutions of the form
\begin{equation}
\xi=\xi_0 t^{\alpha}
\end{equation}
(or an analogous law for $L$),
\begin{equation}
v=v_0 t^{\beta}\,,
\end{equation}
and
\begin{equation}
Q=Q_0 t^{\gamma}\,.
\end{equation}
Note that causality implies $\alpha\leq1$ and the finite speed of light implies $\beta\leq0$. Furthermore, our discussion of loop solutions shows that $v\to1$ is not a physically allowed solution for these networks.


\section{No charge losses}

In this section we assume that there are no macroscopic charge losses. (The case with charge losses will be discussed in the following section.) We will separately consider the cases with and without energy losses due 
to loop production.

Whether or not we have loop production, the evolution equation for the macroscopic charge $Q$ is given by Eq. (\ref{macro30}), and we can start by studying this. There is a trivial but unphysical (refer to discussion after
Eqn. (\ref{macro1})) solution if $k=0$, with $\xi\propto L\propto a$ and $v\propto Q\propto a^{-2}$, which can therefore 
be ignored. In the realistic case $k\neq0$ there can in principle be two kinds of solutions:
\begin{itemize}
\item Decaying charge solutions, with
\begin{equation}
\gamma=-2\lambda\,,\quad \beta<\alpha-1\,;
\end{equation}
for these solutions not only does the charge decay (as $Q\propto a^{-2}$) but velocity will necessarily decay as well.
\item standard solutions with
\begin{equation}
\beta=\alpha-1\,,\quad \gamma=-2\lambda+2k\frac{v_0}{\xi_0}\,;
\end{equation}
here we have used the term `standard' referring to the fact that linear scaling solution (with $\alpha=1$ and $\beta=0$) is of this form, although \textit{a priori} there is no guarantee that this solution will exist with a constant (non-zero) charge. Also note that in this branch of solutions we may at least in principle have growing, constant, or decaying $Q$.
\end{itemize}

We can now study the entire system of equations in the cases with and without energy losses.

\subsection{Without energy losses}

In this case we obtain the following three scaling relations:

\begin{itemize}

\item For slow expansion rates, $\lambda<2/3$,
\begin{equation}\label{slowfirst}
\alpha=\frac{3}{2}\lambda < 1\,,\quad \xi_0=\frac{k v_0}{\lambda}
\end{equation}
\begin{equation}
\beta=\alpha-1<0
\end{equation}
\begin{equation}
\gamma=0\,,\quad Q_0=\left(1+\frac{2s}{k}\right)^{-1}\,
\end{equation}
and
\begin{equation}
\frac{\rho_s}{\rho_{crit}}\propto\frac{\rho}{\rho_{crit}}\propto t^{2-3\lambda}\,;
\end{equation}
here we have a constant charge, which gradually slows the strings (making $v\to0$), although the evolution is still faster than conformal stretching (which corresponds to $\alpha=\lambda$). As a consequence, both the energy density in the strings and the total energy density in the network grow relative to that of the cosmological background.

\item For $\lambda=2/3$, corresponding to the matter-dominated era,
\begin{equation}
\alpha=1\,,\quad \xi_0=\frac{3}{2}k v_0
\end{equation}
\begin{equation}
\beta=0\,,\quad v_0^2=\frac{1}{2}\left[1-Q_0\left(1+\frac{2s}{k}\right)\right]
\end{equation}
\begin{equation}
\gamma=0
\end{equation}
and
\begin{equation}
\frac{\rho_s}{\rho_{crit}}=\frac{1}{1+Q_0}\frac{\rho}{\rho_{crit}}=\frac{16\pi}{3k^2}\left[1-Q_0\left(1+\frac{2s}{k}\right)\right]^{-1}G\mu\,;
\end{equation}
in this case the macroscopic charge is still a constant, but the additional dilution caused by the faster expansion rate is enough to ensure that the energy density of the network is a constant fraction of the background one. In other words, for this particular expansion rate we have a generalised linear scaling solution, with $\xi$ (and $L$) growing as fast as allowed by causality, in which the RMS velocity and the macroscopic charge are constant (and larger charges leading to smaller velocities, as was seen in the loop solutions).

\item For fast expansion rates, $\lambda>2/3$,
\begin{equation}
\alpha=1\,,\quad \xi_0^2=\frac{k^2}{4\lambda (1-\lambda)}
\end{equation}
\begin{equation}
\beta=0\,,\quad v_0^2=\frac{1-\lambda}{\lambda}
\end{equation}
\begin{equation}
\gamma=4-6\lambda<0
\end{equation}
and
\begin{equation}\label{fastlast}
\frac{\rho_s}{\rho_{crit}}=\frac{\rho}{\rho_{crit}}=\frac{32\pi}{3k^2}\frac{1-\lambda}{\lambda}G\mu\,;
\end{equation}
here the additional Hubble damping makes the macroscopic charge decay, and we therefore end up with the solution for ordinary Nambu-Goto strings \cite{KIB,MS1,MS2}.

\end{itemize}

\begin{figure}
\includegraphics[width=3.3in,keepaspectratio]{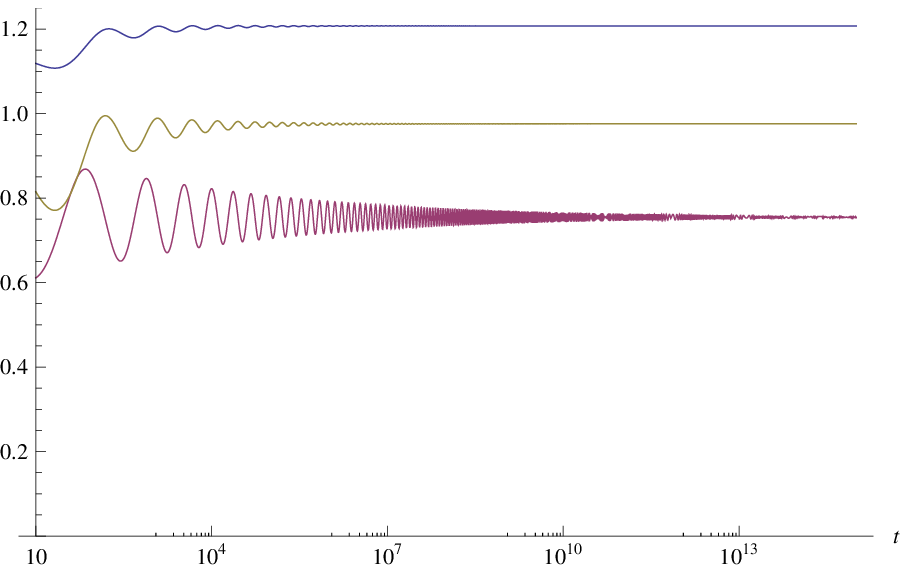}
\includegraphics[width=3.3in,keepaspectratio]{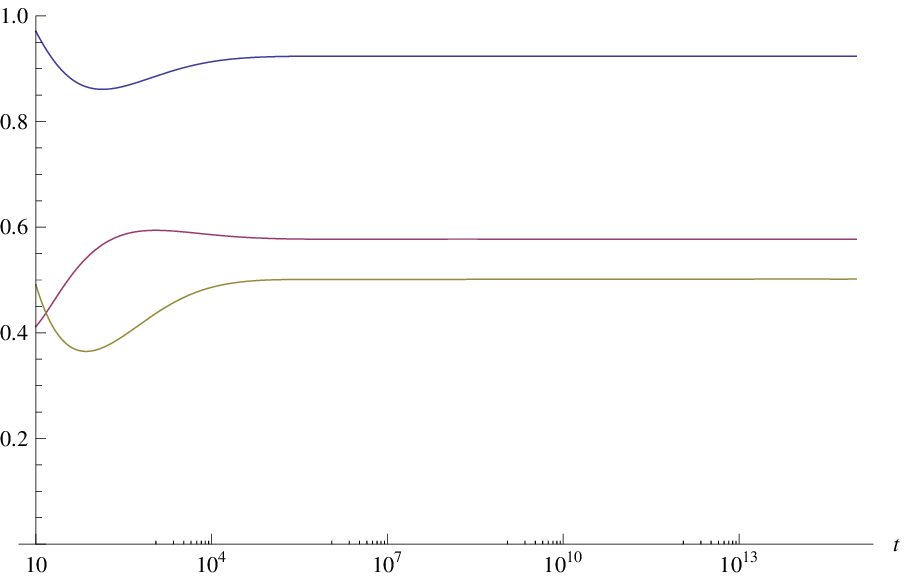}
\caption{\label{fig02}Illustrating the two scaling solutions in the case without energy losses. The plots show the relevant macroscopic quantities divided by the time dependencies inferred in this section's analysis, so that asymptotically constant values imply convergence to the expected scaling solution. {\bf Top panel:} Case $\lambda=1/2$; from top to bottom the lines depict $\xi/t^{3/4}$, $Q$ and $vt^{1/4}$. {\bf Bottom panel:} Case $\lambda=3/4$; from top to bottom the lines depict $\xi/t$, $v$ and $Qt^{1/2}$. In both cases we have assumed $k=0.8$ and $s=0.01$.}
\end{figure}

Therefore, which of these solutions is the attractor for the network's evolution depends on the universe's expansion rate, which is parametrized by $\lambda$. (In principle, one also finds a fourth solution which has $v_0=1$, but, according to our discussion in the loops section, this is unphysical for charged strings.) By numerically solving the evolution equations one can confirm that the above solutions are indeed the attractors in the relevant parameter ranges; Fig.~\ref{fig02} shows examples for the two scaling regimes.

\subsection{With energy losses}

The loss of energy through loop production is described by a new term in the evolution equation for $\dot{\xi}$ (and, consequently, one for $\dot{L}$), which, using standard arguments \cite{KIB,MS1,MS2}, one can write as
\begin{equation}
2\frac{\dot{\xi}}{\xi}=...+\frac{c v}{\xi}\,. 
\end{equation}
This introduces the parameter $c$ quantifying the efficiency of producing loops.
Similarly, using Eq. (\ref{consisteq}) and assuming there is no charge loss, we find the equivalent term for $\dot{L}$,
\begin{equation}
2\frac{\dot{L}}{L}=...+\frac{c v}{L}\frac{1}{\sqrt{1+Q}}\,.
\end{equation}

The analysis of scaling solutions can now be repeated, and one finds solutions that generalise the above ones by including an additional dependency on the loop chopping efficiency $c$:

\begin{itemize}

\item For slow expansion rates, $\lambda<2/(3+c/k)$,
\begin{equation}
\alpha=\frac{1}{2}\left(3+\frac{c}{k}\right)\lambda < 1\,,\quad \xi_0=\frac{k v_0}{\lambda}
\end{equation}
\begin{equation}
\beta=\alpha-1<0
\end{equation}
\begin{equation}
\gamma=0\,,\quad Q_0=\left(1+\frac{2s}{k}\right)^{-1}\,
\end{equation}
and
\begin{equation}
\frac{\rho_s}{\rho_{crit}}\propto\frac{\rho}{\rho_{crit}}\propto t^{2-(3+c/k)\lambda}\,;
\end{equation}
loop production is an additional energy loss mechanism, and therefore the correlation length now grows faster than in the $c=0$ case. Similarly, string velocities decrease more slowly, and the network's density grows more slowly relative to the background one.

\item For an intermediate expansion rate, $\lambda=2/(3+c/k)$,
\begin{equation}
\alpha=1\,,\quad \xi_0=\frac{1}{2} v_0\left(3k+c\right)
\end{equation}
\begin{equation}
\beta=0\,,\quad v_0^2=\frac{1}{2}\left[1-Q_0\left(1+\frac{2s}{k}\right)\right]
\end{equation}
\begin{equation}
\gamma=0
\end{equation}
and
\begin{equation}
\frac{\rho_s}{\rho_{crit}}=\frac{1}{1+Q_0}\frac{\rho}{\rho_{crit}}=\frac{16\pi}{3k^2}\left[1-Q_0\left(1+\frac{2s}{k}\right)\right]^{-1}G\mu\,;
\end{equation}
now the expansion rate for which this solution occurs decreases, with the smaller expansion rate being compensated, for the purposes of the dilution of the network's energy density, by the process of loop production. (For $c=0$ this solution exists for the matter era, and $c=k$ will make it occur in the radiation era.) Interestingly, the ratio of the string and background energies is exactly the same as before---in other words, it does not depend on the value of $c$. 

\item For fast expansion rates, $\lambda>2/(3+c/k)$,
\begin{equation}
\alpha=1\,,\quad \xi_0^2=\frac{k(k+c)}{4\lambda (1-\lambda)}
\end{equation}
\begin{equation}
\beta=0\,,\quad v_0^2=\frac{1-\lambda}{\lambda}\frac{k}{k+c}
\end{equation}
\begin{equation}
\gamma=\frac{4}{k+c}\left[k-\frac{\lambda}{2}(3k+c)\right]<0
\end{equation}
and
\begin{equation}
\frac{\rho_s}{\rho_{crit}}=\frac{\rho}{\rho_{crit}}=\frac{32\pi}{3k(k+c)}\frac{1-\lambda}{\lambda}G\mu\,;
\end{equation}
this is exactly the VOS linear scaling solution for Nambu-Goto strings \cite{MS1,MS2,MS3,MS4}, with an added prediction of a particular decay law for the charge (which depends on the cosmological expansion rate). 

\end{itemize}

\begin{figure}
\includegraphics[width=3.3in,keepaspectratio]{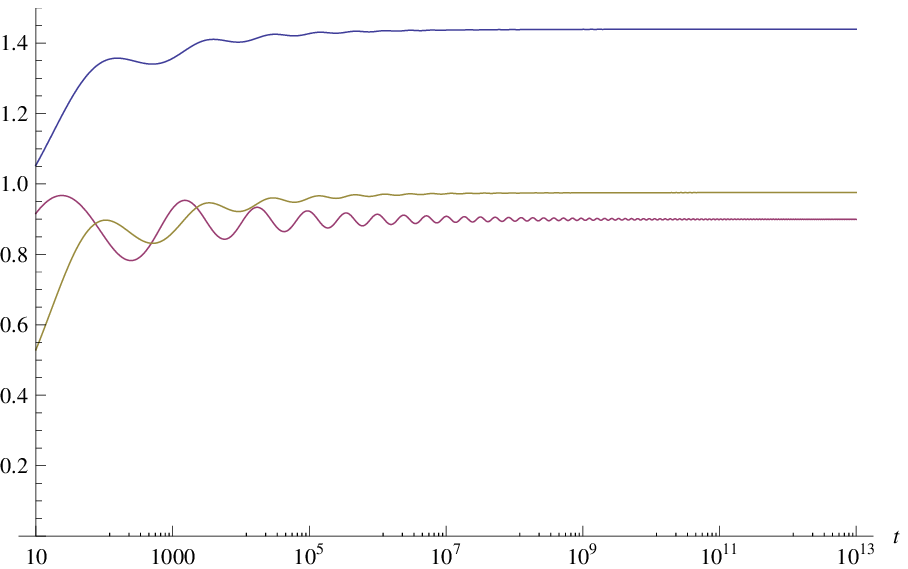}
\includegraphics[width=3.3in,keepaspectratio]{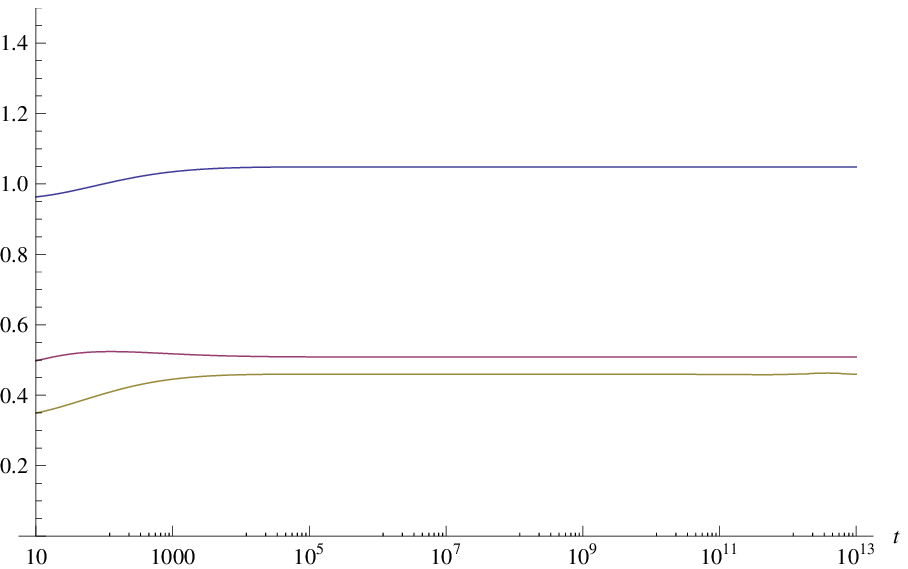}
\caption{\label{fig03}Illustrating the two scaling solutions in the case with energy losses. The plots show the relevant macroscopic quantities divided by the time dependencies inferred in this section's analysis, so that asymptotically constant values imply convergence to the expected scaling solution. {\bf Top panel:} Case $\lambda=1/2$; from top to bottom the lines depict $\xi/t^{0.822...}$, $Q$ and $vt^{0.178...}$. {\bf Bottom panel:} Case $\lambda=3/4$; from top to bottom the lines depict $\xi/t$, $v$ and $Qt^{0.723...}$. In both cases we have assumed $k=0.8$, $c=0.23$ and $s=0.01$.}
\end{figure}

Fig.~\ref{fig03} illustrates the two branches of the solution. It can be trivially seen that in the limit of no loop production, $c\to0$, these solutions are the same as in the case without energy losses, Eqns. (\ref{slowfirst})-(\ref{fastlast}). As before, there is a fourth solution with $v=1$ which is not allowed for charged strings. The analysis so far also highlights the point that the parameter $c$ is important for determining whether charge survives on the strings or decays. This motivates a discussion of possible charge losses, which will be done in the next section.

\section{Macroscopic charge losses}

In the previous section, in order to derive the energy loss term in the evolution equation for $L$, it was assumed that there was no charge loss. Now this assumption will be dropped, and the resulting scaling relations analysed. This section is somewhat more phenomenological than the rest of the article, since it will rely on simplifying assumptions for how the charge may be lost, but our goal is simply to develop an intuitive picture for the possible role of charge loss mechanisms on the evolution of the network.

We will again assume an energy loss term of the form
\begin{equation}
\frac{\dot E_s}{E_s}=-c\frac{v}{\xi}\,,\quad 2\frac{\dot{\xi}}{\xi}=...+c\frac{v}{\xi}
\end{equation}
but this time let us say that $L$ gets a different (in general) loss term
\begin{equation}
\frac{\dot E}{E}=-fc\frac{v}{L}\,,\quad 2\frac{\dot L}{L}=...+fc\frac{v}{L} \,,
\end{equation}
where $f$ is an arbitrary function, possibly of the velocity, charge and correlation length. We are therefore assuming that any such charge losses are related to the network's intercommuting and loop production mechanisms.

In the previous section, we assumed there was no charge loss, implicitly using
\begin{equation}\label{fchoice}
f(v,Q,\xi)= \frac{1}{\sqrt{1+Q}}\,.
\end{equation}
Instead, we will now leave $f$ free and obtain the evolution equation for $Q$ from the above assumptions together with Eq. (\ref{consisteq}). The result is
\begin{equation}
\dot{Q}=2Q(\frac{kv}{\xi}-H)+\frac{cv}{\xi}(1+Q)\left(1-f\sqrt{1+Q}\right)\,,
\end{equation}
or an analogous equation in terms of $L$. With the choice (\ref{fchoice}) for $f$ we trivially recover the results of the previous sections. We can now look for scaling relations as before, checking how the results depend on the choice of $f$.

We start by noting that, in the standard VOS-model case (without a charge), we would not expect charge to be created. Therefore, $\dot{Q}=0$ when $Q=0$, implying
\begin{equation}
2\frac{cv}{\xi}(1-f)=0\,.
\end{equation}
It then follows that, if the function $f$ is constant, it has to be equal to unity. Otherwise it must be dependent on the charge (and possibly other quantities), and become unity when the charge drops to zero,
\begin{equation}
f=f(Q,v,\xi)\implies f(0,v,\xi)=1\,,
\end{equation}
as was the case for the function (\ref{fchoice}) we used in the case without charge loss.

Let us therefore consider the $f=1$ case. The generalised scaling laws now become:

\begin{itemize}

\item For slow expansion rates, $\lambda<(2k-Wc)/[3k+(1-W)c]$,
\begin{equation}
\alpha=\frac{3}{2}\left[\frac{1+(1-W)c/3k}{1-Wc/2k}\right]\lambda < 1\,,\quad \xi_0=\frac{v_0}{\lambda}(k-Wc/2)
\end{equation}
\begin{equation}
\beta=\alpha-1<0
\end{equation}
\begin{equation}
\gamma=0\,,\quad Q_0=\left(1+\frac{2s}{k}\right)^{-1}\,
\end{equation}
and naturally
\begin{equation}
\frac{\rho_s}{\rho_{crit}}\propto\frac{\rho}{\rho_{crit}}\propto t^{2(1-\alpha)}\,
\end{equation}
(where for simplicity we have kept $\alpha$ in the last expression); we have also introduced
\begin{equation}
W=(1+Q_0^{-1})\left(\sqrt{1+Q_0}-1\right)\,,
\end{equation}
which is always positive and behaves as $W=0.5(1+Q_0)$ in the limit $Q_0\to0$ and as $W\propto \sqrt{Q_0}$ for $Q_0\to\infty$. Notice that the scaling exponent for $\xi$ now has an explicit dependence on the charge, which was not the case without charge losses.

As one would expect, increasing the scaling value of the charge pushes the value of the maximal expansion rate where this regime holds to larger values, whereas the scaling exponent $\alpha$ decreases. Similarly, string velocities decrease faster, and the network's density  grows faster relative to the background one.

In principle, as one makes $W$ progressively larger, the scaling exponent $\alpha$ becomes closer to $\lambda$, which corresponds to the conformal stretching case. However, in practice we do not expect this to occur, since it is clear from its definition that $W$ should be a small parameter in realistic (cosmological) defect networks.

\item For an intermediate expansion rate, $\lambda=(2k-Wc)/[3k+(1-W)c]$,
\begin{equation}
\alpha=1\,,\quad \xi_0=\frac{1}{2} v_0\left[3k+(1-W)c\right]
\end{equation}
\begin{equation}
\beta=0\,,\quad v_0^2=\frac{1}{2-Wc/k}\left[1-Q_0\left(1+\frac{2s}{k}\right)\right]
\end{equation}
\begin{equation}
\gamma=0
\end{equation}
and
\begin{equation}
\frac{\rho_s}{\rho_{crit}}=\frac{1}{1+Q_0}\frac{\rho}{\rho_{crit}}=\frac{16\pi}{3k^2}\left[1-Q_0\left(1+\frac{2s}{k}\right)\right]^{-1}G\mu\,.
\end{equation}
As in the previous solution there is now an explicit dependence on the amount of charge, and the expansion rate for which this solution occurs increases with the charge. This is to be expected: a larger charge makes scaling harder, requiring more energy losses (from the damping due to the Hubble expansion, or from losses due to loop production) to counteract it. In fact the expansion rate for which this solution exists would approach $\lambda=1$ as the scaling value of the charge $Q_0$ becomes arbitrarily large, although as we already pointed out we do not expect this to occur in practice. The ratio of the string and background energies is still exactly the same as before---in other words, it does not depend on the value of $c$ or $W$. 

\item For fast expansion rates, $\lambda>(2k-W_0c)/[3k+(1-W_0)c]$,
\begin{equation}
\alpha=1\,,\quad \xi_0^2=\frac{k(k+c)}{4\lambda (1-\lambda)}
\end{equation}
\begin{equation}
\beta=0\,,\quad v_0^2=\frac{1-\lambda}{\lambda}\frac{k}{k+c}
\end{equation}
\begin{equation}
\gamma=\frac{4}{k+c}\left[k-\frac{W_0c}{2}-\frac{\lambda}{2}[3k+(1-W_0)c]\right]<0
\end{equation}
and
\begin{equation}
\frac{\rho_s}{\rho_{crit}}=\frac{\rho}{\rho_{crit}}=\frac{32\pi}{3k(k+c)}\frac{1-\lambda}{\lambda}G\mu\,;
\end{equation}
which is again the VOS linear scaling solution for Nambu-Goto strings, now with a faster decay law for the charge (which is obvious since we have explicit charge losses). For this solution we used the notation $W_0$ to indicate the value of $W$ in the limit $Q_0\to0$, that is $W=1/2$; the reason for this choice will become clear below.
\end{itemize}

Again it is easy to check that if we set $c=0$ and/or $W=0$ we recover the results of the previous section. However, note
that the transition between the second and third solutions will now depend on the amount of charge loss. The expansion rate of the second solution coincides with the minimum expansion rate of the third solution for
\begin{equation}
W=W_0=\frac{1}{2}\,,
\end{equation}
which corresponds to the $Q_0=0$ limit.

Finally, it is interesting to discuss what happens in the more general case where $f\neq1$. In the absence of compelling arguments suggesting a particular form for $f$ (other than the case without charge losses which we studied in the previous section) we will consider a linearised form, that is
\begin{equation}
f(Q)=1+wQ\,,
\end{equation}
for $w$ real with $|w|<1$.
The rationale for this is that in realistic networks in cosmological contexts the charges are 
likely to correspond to a small fraction of the overall energy density of the network.

Repeating the analysis we find that the above solutions still hold, provided that one extends the definition of the parameter $W$ to
\begin{equation}
W=(1+Q_0^{-1})\left[(1+w Q_0)\sqrt{1+Q_0}-1\right]\,.
\end{equation}
In the limit $Q_0\to0$ this now behaves as
\begin{equation}
W=\left(\frac{1}{2}+w \right)(1+Q_0)\,;
\end{equation}
we trivially recover the behaviour in the $w=0$ (constant) case, but we also see that in this limit $W$ will vanish if $w=-1/2$. This is interesting because that choice of $w$ corresponds to the linearised version of 
$f=1/\sqrt{1+Q}$,
which as we argued is the case of no charge losses. This shows that the above analysis is self-consistent.

\begin{figure}
\includegraphics[width=3.3in,keepaspectratio]{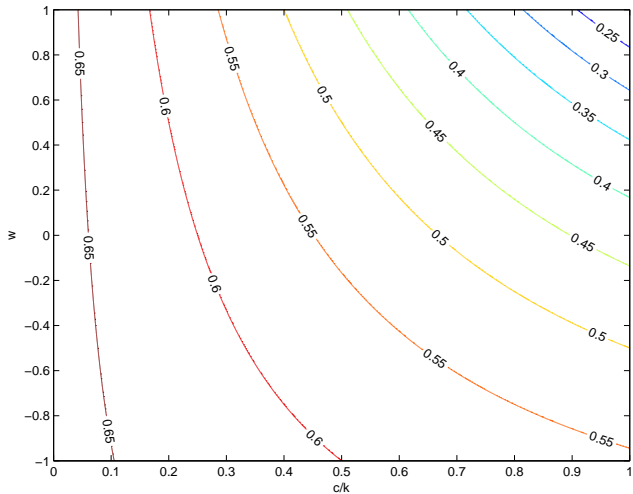}
\caption{\label{critical}The value of the expansion rate $\lambda$ at which the attractor for the network's evolution changes from a fixed charge solution to the standard linear scaling one, as a function of the parameters $w$ and $c/k$.}
\end{figure}

\section{Discussion and conclusions}

We have extended the velocity-dependent one-scale string evolution model to the case where there is a conserved microscopic charge on the string worldsheet. We find that there are two possible regimes for the evolution of the network. When the expansion rate of the universe is fast enough the macroscopic charge will decay and the attractor solution is the standard linear scaling regime, where the network is losing energy as fast as is allowed by causality. When the expansion rate is relatively slow, the attractor solution has a constant macroscopic charge, with the defect velocities decreasing and the network correlation length growing more slowly than allowed by causality. However, even in this case, the network does not generically frustrate: the correlation length evolves more slowly than $\xi\propto t$ but faster than $\xi\propto a$ (conformal stretching); only in the limit of arbitrarily large charge would the conformal stretching behaviour be reached.

The rate of expansion at which the transition from charge domination to linear scaling occurs will depend on the dynamical processes acting on the network. A larger charge makes scaling harder, but cosmological expansion provides a damping term which tends to dilute the charge and overcome its influence on the dynamics. Further energy loss mechanisms facilitate scaling. Specifically, this critical expansion rate is given, in terms of the model parameters identified in the previous sections, by
\begin{equation}
\lambda=\frac{2}{3}\,\frac{1-(0.5+w)c/2k}{1+(0.5-w)c/3k}\,.
\end{equation}
Without any losses the exponent is $2/3$ (corresponding to the matter-dominated era), and its dependence on the model parameters is illustrated in Fig.~\ref{critical}.

While our study was purely analytical, we should note that some numerical work already exists which, at least qualitatively, is in agreement with our results. Specifically, the aforementioned results of \cite{BATTYE1,BATTYE2}, which suggest violations of scaling, rely on simulations which are nominally done in Minkowski space, although they have some effective numerical damping which probably mimics a fairly small (but non-zero) expansion rate. This therefore agrees with our results for low expansion rates, if one assumes that there are some effective charge losses. In that case our results predict that the scaling exponent for $\xi$ will decrease as the charge increases. However, we would further suggest that in a cosmological setting (that is, with  a faster expansion rate), the network may still scale.

Interestingly, something like that kind of behaviour seems to emerge in more recent work \cite{BATTYEflaw}, in which apparent scaling deviations in Minkowski space (for a different model, again in two dimensions) are in fact eliminated by increasing the amount of damping. A more quantitative comparison between our analytic solutions and these numerical results would certainly be instructive but is not possible at this stage: it would require knowledge of the behaviour of the defect velocities in the simulations, and that is not provided in any of these works \cite{BATTYE1,BATTYE2,BATTYEflaw}.

In any case, our analysis highlights the point that results derived in Minkowski space may not necessarily apply to the case of an expanding universe; this has already been discussed, for example for the case of wiggly cosmic strings in \cite{Fractal}. Finally, our results are also relevant for the case of superconducting strings and vortons, but this case is left for future work.

\section*{Acknowledgements}
This work was done in the context of the research grant PTDC/FIS/111725/2009 from FCT, Portugal. Useful discussions with Paul Shellard on this subject are gratefully acknowledged. 
A.A. was supported by a Marie Curie IEF Fellowship at the the University of Nottingham.
The work of CM is funded by a Ci\^encia2007 Research Contract, funded by FCT/MCTES (Portugal) and POPH/FSE (EC). MO and CM also acknowledge the hospitality of the University of Nottingham, where this work was completed.

\bibliography{charged}

\begin{thebibliography}{44}
\expandafter\ifx\csname natexlab\endcsname\relax\def\natexlab#1{#1}\fi
\expandafter\ifx\csname bibnamefont\endcsname\relax
  \def\bibnamefont#1{#1}\fi
\expandafter\ifx\csname bibfnamefont\endcsname\relax
  \def\bibfnamefont#1{#1}\fi
\expandafter\ifx\csname citenamefont\endcsname\relax
  \def\citenamefont#1{#1}\fi
\expandafter\ifx\csname url\endcsname\relax
  \def\url#1{\texttt{#1}}\fi
\expandafter\ifx\csname urlprefix\endcsname\relax\def\urlprefix{URL }\fi
\providecommand{\bibinfo}[2]{#2}
\providecommand{\eprint}[2][]{\url{#2}}

\bibitem[{\citenamefont{Vilenkin and Shellard}(1994)}]{VSH}
\bibinfo{author}{\bibfnamefont{A.}~\bibnamefont{Vilenkin}} \bibnamefont{and}
  \bibinfo{author}{\bibfnamefont{E.~P.~S.} \bibnamefont{Shellard}},
  \emph{\bibinfo{title}{Cosmic Strings and other Topological Defects}}
  (\bibinfo{publisher}{Cambridge University Press},
  \bibinfo{address}{Cambridge, U.K.}, \bibinfo{year}{1994}).

\bibitem[{\citenamefont{Hindmarsh and Kibble}(1995)}]{HindKib}
\bibinfo{author}{\bibfnamefont{M.~B.} \bibnamefont{Hindmarsh}}
  \bibnamefont{and} \bibinfo{author}{\bibfnamefont{T.~W.~B.}
  \bibnamefont{Kibble}}, \bibinfo{journal}{Rept.Prog.Phys.}
  \textbf{\bibinfo{volume}{58}}, \bibinfo{pages}{477} (\bibinfo{year}{1995}),
  \eprint{hep-ph/9411342}.

\bibitem[{\citenamefont{Copeland et~al.}(2011)\citenamefont{Copeland, Pogosian,
  and Vachaspati}}]{CopPogVach}
\bibinfo{author}{\bibfnamefont{E.~J.} \bibnamefont{Copeland}},
  \bibinfo{author}{\bibfnamefont{L.}~\bibnamefont{Pogosian}}, \bibnamefont{and}
  \bibinfo{author}{\bibfnamefont{T.}~\bibnamefont{Vachaspati}},
  \bibinfo{journal}{Class.Quant.Grav.} \textbf{\bibinfo{volume}{28}},
  \bibinfo{pages}{204009} (\bibinfo{year}{2011}), \eprint{1105.0207}.

\bibitem[{\citenamefont{Kibble}(2004)}]{KibbleRev}
\bibinfo{author}{\bibfnamefont{T.~W.~B.} \bibnamefont{Kibble}}
  (\bibinfo{year}{2004}), \eprint{astro-ph/0410073}.

\bibitem[{\citenamefont{Jeannerot et~al.}(2003)\citenamefont{Jeannerot, Rocher,
  and Sakellariadou}}]{JeRoSak}
\bibinfo{author}{\bibfnamefont{R.}~\bibnamefont{Jeannerot}},
  \bibinfo{author}{\bibfnamefont{J.}~\bibnamefont{Rocher}}, \bibnamefont{and}
  \bibinfo{author}{\bibfnamefont{M.}~\bibnamefont{Sakellariadou}},
  \bibinfo{journal}{Phys. Rev.} \textbf{\bibinfo{volume}{D68}},
  \bibinfo{pages}{103514} (\bibinfo{year}{2003}), \eprint{hep-ph/0308134}.

\bibitem[{\citenamefont{Burgess et~al.}(2001)\citenamefont{Burgess, Majumdar,
  Quevedo, Rajesh, and Zhang}}]{BMNQRZ}
\bibinfo{author}{\bibfnamefont{C.~P.} \bibnamefont{Burgess}},
  \bibinfo{author}{\bibfnamefont{M.}~\bibnamefont{Majumdar}},
  \bibinfo{author}{\bibfnamefont{F.}~\bibnamefont{Quevedo}},
  \bibinfo{author}{\bibfnamefont{G.}~\bibnamefont{Rajesh}}, \bibnamefont{and}
  \bibinfo{author}{\bibfnamefont{R.-J.} \bibnamefont{Zhang}},
  \bibinfo{journal}{JHEP} \textbf{\bibinfo{volume}{0107}}, \bibinfo{pages}{047}
  (\bibinfo{year}{2001}), \eprint{hep-th/0105204}.

\bibitem[{\citenamefont{Sarangi and Tye}(2002)}]{SarTye}
\bibinfo{author}{\bibfnamefont{S.}~\bibnamefont{Sarangi}} \bibnamefont{and}
  \bibinfo{author}{\bibfnamefont{S.~H.~H.} \bibnamefont{Tye}},
  \bibinfo{journal}{Phys. Lett.} \textbf{\bibinfo{volume}{B536}},
  \bibinfo{pages}{185} (\bibinfo{year}{2002}), \eprint{hep-th/0204074}.

\bibitem[{\citenamefont{Dvali and Vilenkin}(2004)}]{DvalVil}
\bibinfo{author}{\bibfnamefont{G.}~\bibnamefont{Dvali}} \bibnamefont{and}
  \bibinfo{author}{\bibfnamefont{A.}~\bibnamefont{Vilenkin}},
  \bibinfo{journal}{JCAP} \textbf{\bibinfo{volume}{0403}}, \bibinfo{pages}{010}
  (\bibinfo{year}{2004}), \eprint{hep-th/0312007}.

\bibitem[{\citenamefont{Copeland et~al.}(2004)\citenamefont{Copeland, Myers,
  and Polchinski}}]{CopMyePolch}
\bibinfo{author}{\bibfnamefont{E.~J.} \bibnamefont{Copeland}},
  \bibinfo{author}{\bibfnamefont{R.~C.} \bibnamefont{Myers}}, \bibnamefont{and}
  \bibinfo{author}{\bibfnamefont{J.}~\bibnamefont{Polchinski}},
  \bibinfo{journal}{JHEP} \textbf{\bibinfo{volume}{06}}, \bibinfo{pages}{013}
  (\bibinfo{year}{2004}), \eprint{hep-th/0312067}.

\bibitem[{\citenamefont{Pourtsidou et~al.}(2011)\citenamefont{Pourtsidou,
  Avgoustidis, Copeland, Pogosian, and Steer}}]{PACPS}
\bibinfo{author}{\bibfnamefont{A.}~\bibnamefont{Pourtsidou}},
  \bibinfo{author}{\bibfnamefont{A.}~\bibnamefont{Avgoustidis}},
  \bibinfo{author}{\bibfnamefont{E.~J.} \bibnamefont{Copeland}},
  \bibinfo{author}{\bibfnamefont{L.}~\bibnamefont{Pogosian}}, \bibnamefont{and}
  \bibinfo{author}{\bibfnamefont{D.~A.} \bibnamefont{Steer}},
  \bibinfo{journal}{Phys.Rev.} \textbf{\bibinfo{volume}{D83}},
  \bibinfo{pages}{063525} (\bibinfo{year}{2011}), \eprint{1012.5014}.

\bibitem[{\citenamefont{Avgoustidis et~al.}(2011)\citenamefont{Avgoustidis,
  Copeland, Moss, Pogosian, Pourtsidou et~al.}}]{ACMPPS}
\bibinfo{author}{\bibfnamefont{A.}~\bibnamefont{Avgoustidis}},
  \bibinfo{author}{\bibfnamefont{E.~J.} \bibnamefont{Copeland}},
  \bibinfo{author}{\bibfnamefont{A.}~\bibnamefont{Moss}},
  \bibinfo{author}{\bibfnamefont{L.}~\bibnamefont{Pogosian}},
  \bibinfo{author}{\bibfnamefont{A.}~\bibnamefont{Pourtsidou}},
  \bibnamefont{et~al.}, \bibinfo{journal}{Phys.Rev.Lett.}
  \textbf{\bibinfo{volume}{107}}, \bibinfo{pages}{121301}
  (\bibinfo{year}{2011}), \eprint{1105.6198}.

\bibitem[{\citenamefont{Copeland and Saffin}(2005)}]{CopSaf}
\bibinfo{author}{\bibfnamefont{E.~J.} \bibnamefont{Copeland}} \bibnamefont{and}
  \bibinfo{author}{\bibfnamefont{P.~M.} \bibnamefont{Saffin}},
  \bibinfo{journal}{JHEP} \textbf{\bibinfo{volume}{0511}}, \bibinfo{pages}{023}
  (\bibinfo{year}{2005}), \eprint{hep-th/0505110}.

\bibitem[{\citenamefont{Urrestilla and Vilenkin}(2008)}]{UrrVil}
\bibinfo{author}{\bibfnamefont{J.}~\bibnamefont{Urrestilla}} \bibnamefont{and}
  \bibinfo{author}{\bibfnamefont{A.}~\bibnamefont{Vilenkin}},
  \bibinfo{journal}{JHEP} \textbf{\bibinfo{volume}{0802}}, \bibinfo{pages}{037}
  (\bibinfo{year}{2008}), \eprint{0712.1146}.

\bibitem[{\citenamefont{Hindmarsh and Saffin}(2006)}]{HindSaf}
\bibinfo{author}{\bibfnamefont{M.}~\bibnamefont{Hindmarsh}} \bibnamefont{and}
  \bibinfo{author}{\bibfnamefont{P.~M.} \bibnamefont{Saffin}},
  \bibinfo{journal}{JHEP} \textbf{\bibinfo{volume}{0608}}, \bibinfo{pages}{066}
  (\bibinfo{year}{2006}), \eprint{hep-th/0605014}.

\bibitem[{\citenamefont{Sakellariadou and Stoica}(2008)}]{SakSto}
\bibinfo{author}{\bibfnamefont{M.}~\bibnamefont{Sakellariadou}}
  \bibnamefont{and} \bibinfo{author}{\bibfnamefont{H.}~\bibnamefont{Stoica}},
  \bibinfo{journal}{JCAP} \textbf{\bibinfo{volume}{0808}}, \bibinfo{pages}{038}
  (\bibinfo{year}{2008}), \eprint{0806.3219}.

\bibitem[{\citenamefont{Tye et~al.}(2005)\citenamefont{Tye, Wasserman, and
  Wyman}}]{TWW}
\bibinfo{author}{\bibfnamefont{S.-H.} \bibnamefont{Tye}},
  \bibinfo{author}{\bibfnamefont{I.}~\bibnamefont{Wasserman}},
  \bibnamefont{and} \bibinfo{author}{\bibfnamefont{M.}~\bibnamefont{Wyman}},
  \bibinfo{journal}{Phys.Rev.} \textbf{\bibinfo{volume}{D71}},
  \bibinfo{pages}{103508} (\bibinfo{year}{2005}), \eprint{astro-ph/0503506}.

\bibitem[{\citenamefont{Avgoustidis and Shellard}(2008)}]{AvgShell}
\bibinfo{author}{\bibfnamefont{A.}~\bibnamefont{Avgoustidis}} \bibnamefont{and}
  \bibinfo{author}{\bibfnamefont{E.~P.~S.} \bibnamefont{Shellard}},
  \bibinfo{journal}{Phys.Rev.} \textbf{\bibinfo{volume}{D78}},
  \bibinfo{pages}{103510} (\bibinfo{year}{2008}), \eprint{0705.3395}.

\bibitem[{\citenamefont{Avgoustidis and Copeland}(2010)}]{AvgCop}
\bibinfo{author}{\bibfnamefont{A.}~\bibnamefont{Avgoustidis}} \bibnamefont{and}
  \bibinfo{author}{\bibfnamefont{E.~J.} \bibnamefont{Copeland}},
  \bibinfo{journal}{Phys.Rev.} \textbf{\bibinfo{volume}{D81}},
  \bibinfo{pages}{063517} (\bibinfo{year}{2010}), \eprint{0912.4004}.

\bibitem[{\citenamefont{Polchinski}(2004)}]{PolchIntro}
\bibinfo{author}{\bibfnamefont{J.}~\bibnamefont{Polchinski}}, pp.
  \bibinfo{pages}{229--253} (\bibinfo{year}{2004}), \eprint{hep-th/0412244}.

\bibitem[{\citenamefont{Witten}(1985)}]{Witten}
\bibinfo{author}{\bibfnamefont{E.}~\bibnamefont{Witten}},
  \bibinfo{journal}{Nucl.Phys.} \textbf{\bibinfo{volume}{B249}},
  \bibinfo{pages}{557} (\bibinfo{year}{1985}).

\bibitem[{\citenamefont{Callan and Harvey}(1985)}]{CalJef}
\bibinfo{author}{\bibfnamefont{C.~G.} \bibnamefont{Callan}} \bibnamefont{and}
  \bibinfo{author}{\bibfnamefont{J.~A.} \bibnamefont{Harvey}},
  \bibinfo{journal}{Nucl.Phys.} \textbf{\bibinfo{volume}{B250}},
  \bibinfo{pages}{427} (\bibinfo{year}{1985}).

\bibitem[{\citenamefont{Davis and Shellard}(1988)}]{Vorton1}
\bibinfo{author}{\bibfnamefont{R.~L.} \bibnamefont{Davis}} \bibnamefont{and}
  \bibinfo{author}{\bibfnamefont{E.~P.~S.} \bibnamefont{Shellard}},
  \bibinfo{journal}{Phys.Lett.} \textbf{\bibinfo{volume}{B209}},
  \bibinfo{pages}{485} (\bibinfo{year}{1988}).

\bibitem[{\citenamefont{Ganoulis and Lazarides}(1989)}]{GanLaz}
\bibinfo{author}{\bibfnamefont{N.}~\bibnamefont{Ganoulis}} \bibnamefont{and}
  \bibinfo{author}{\bibfnamefont{G.}~\bibnamefont{Lazarides}},
  \bibinfo{journal}{Nucl.Phys.} \textbf{\bibinfo{volume}{B316}},
  \bibinfo{pages}{443} (\bibinfo{year}{1989}).

\bibitem[{\citenamefont{Carter and Peter}(1995)}]{CarPet}
\bibinfo{author}{\bibfnamefont{B.}~\bibnamefont{Carter}} \bibnamefont{and}
  \bibinfo{author}{\bibfnamefont{P.}~\bibnamefont{Peter}},
  \bibinfo{journal}{Phys.Rev.} \textbf{\bibinfo{volume}{D52}},
  \bibinfo{pages}{1744} (\bibinfo{year}{1995}), \eprint{hep-ph/9411425}.

\bibitem[{\citenamefont{Brax et~al.}(2006)\citenamefont{Brax, van~de Bruck,
  Davis, and Davis}}]{BvdBDD}
\bibinfo{author}{\bibfnamefont{P.}~\bibnamefont{Brax}},
  \bibinfo{author}{\bibfnamefont{C.}~\bibnamefont{van~de Bruck}},
  \bibinfo{author}{\bibfnamefont{A.}~\bibnamefont{Davis}}, \bibnamefont{and}
  \bibinfo{author}{\bibfnamefont{S.~C.} \bibnamefont{Davis}},
  \bibinfo{journal}{JHEP} \textbf{\bibinfo{volume}{0606}}, \bibinfo{pages}{030}
  (\bibinfo{year}{2006}), \eprint{hep-th/0604198}.

\bibitem[{\citenamefont{Davis and Shellard}(1989)}]{Vorton2}
\bibinfo{author}{\bibfnamefont{R.~L.} \bibnamefont{Davis}} \bibnamefont{and}
  \bibinfo{author}{\bibfnamefont{E.~P.~S.} \bibnamefont{Shellard}},
  \bibinfo{journal}{Nucl.Phys.} \textbf{\bibinfo{volume}{B323}},
  \bibinfo{pages}{209} (\bibinfo{year}{1989}).

\bibitem[{\citenamefont{Brandenberger et~al.}(1996)\citenamefont{Brandenberger,
  Carter, Davis, and Trodden}}]{BCDT}
\bibinfo{author}{\bibfnamefont{R.~H.} \bibnamefont{Brandenberger}},
  \bibinfo{author}{\bibfnamefont{B.}~\bibnamefont{Carter}},
  \bibinfo{author}{\bibfnamefont{A.-C.} \bibnamefont{Davis}}, \bibnamefont{and}
  \bibinfo{author}{\bibfnamefont{M.}~\bibnamefont{Trodden}},
  \bibinfo{journal}{Phys.Rev.} \textbf{\bibinfo{volume}{D54}},
  \bibinfo{pages}{6059} (\bibinfo{year}{1996}), \eprint{hep-ph/9605382}.

\bibitem[{\citenamefont{Martins and Shellard}(1998{\natexlab{a}})}]{SUPERC2}
\bibinfo{author}{\bibfnamefont{C.~J. A.~P.} \bibnamefont{Martins}}
  \bibnamefont{and} \bibinfo{author}{\bibfnamefont{E.~P.~S.}
  \bibnamefont{Shellard}}, \bibinfo{journal}{Phys. Rev.}
  \textbf{\bibinfo{volume}{D57}}, \bibinfo{pages}{7155}
  (\bibinfo{year}{1998}{\natexlab{a}}), \eprint{hep-ph/9804378}.

\bibitem[{\citenamefont{Martins and Shellard}(1998{\natexlab{b}})}]{Vorton3}
\bibinfo{author}{\bibfnamefont{C.~J. A.~P.} \bibnamefont{Martins}}
  \bibnamefont{and} \bibinfo{author}{\bibfnamefont{E.~P.~S.}
  \bibnamefont{Shellard}}, \bibinfo{journal}{Phys.Lett.}
  \textbf{\bibinfo{volume}{B445}}, \bibinfo{pages}{43}
  (\bibinfo{year}{1998}{\natexlab{b}}), \eprint{hep-ph/9806480}.

\bibitem[{\citenamefont{Nunes et~al.}(2011)\citenamefont{Nunes, Avgoustidis,
  Martins, and Urrestilla}}]{NAMU}
\bibinfo{author}{\bibfnamefont{A.~S.} \bibnamefont{Nunes}},
  \bibinfo{author}{\bibfnamefont{A.}~\bibnamefont{Avgoustidis}},
  \bibinfo{author}{\bibfnamefont{C.~J. A.~P.} \bibnamefont{Martins}},
  \bibnamefont{and}
  \bibinfo{author}{\bibfnamefont{J.}~\bibnamefont{Urrestilla}},
  \bibinfo{journal}{Phys.Rev.} \textbf{\bibinfo{volume}{D84}},
  \bibinfo{pages}{063504} (\bibinfo{year}{2011}), \eprint{1107.2008}.

\bibitem[{\citenamefont{Vachaspati and Achucarro}(1991)}]{VachAch}
\bibinfo{author}{\bibfnamefont{T.}~\bibnamefont{Vachaspati}} \bibnamefont{and}
  \bibinfo{author}{\bibfnamefont{A.}~\bibnamefont{Achucarro}},
  \bibinfo{journal}{Phys.Rev.} \textbf{\bibinfo{volume}{D44}},
  \bibinfo{pages}{3067} (\bibinfo{year}{1991}).

\bibitem[{\citenamefont{Martins and Shellard}(1996{\natexlab{a}})}]{MS1}
\bibinfo{author}{\bibfnamefont{C.~J. A.~P.} \bibnamefont{Martins}}
  \bibnamefont{and} \bibinfo{author}{\bibfnamefont{E.~P.~S.}
  \bibnamefont{Shellard}}, \bibinfo{journal}{Phys. Rev.}
  \textbf{\bibinfo{volume}{D53}}, \bibinfo{pages}{575}
  (\bibinfo{year}{1996}{\natexlab{a}}), \eprint{hep-ph/9507335}.

\bibitem[{\citenamefont{Martins and Shellard}(1996{\natexlab{b}})}]{MS2}
\bibinfo{author}{\bibfnamefont{C.~J. A.~P.} \bibnamefont{Martins}}
  \bibnamefont{and} \bibinfo{author}{\bibfnamefont{E.~P.~S.}
  \bibnamefont{Shellard}}, \bibinfo{journal}{Phys. Rev.}
  \textbf{\bibinfo{volume}{D54}}, \bibinfo{pages}{2535}
  (\bibinfo{year}{1996}{\natexlab{b}}),
  \eprint[http://arXiv.org/abs]{hep-ph/9602271}.

\bibitem[{\citenamefont{Martins and Shellard}(2002)}]{MS3}
\bibinfo{author}{\bibfnamefont{C.~J. A.~P.} \bibnamefont{Martins}}
  \bibnamefont{and} \bibinfo{author}{\bibfnamefont{E.~P.~S.}
  \bibnamefont{Shellard}}, \bibinfo{journal}{Phys. Rev.}
  \textbf{\bibinfo{volume}{D65}}, \bibinfo{pages}{043514}
  (\bibinfo{year}{2002}), \eprint[http://arXiv.org/abs]{hep-ph/0003298}.

\bibitem[{\citenamefont{Martins et~al.}(2004)\citenamefont{Martins, Moore, and
  Shellard}}]{MS4}
\bibinfo{author}{\bibfnamefont{C.~J. A.~P.} \bibnamefont{Martins}},
  \bibinfo{author}{\bibfnamefont{J.~N.} \bibnamefont{Moore}}, \bibnamefont{and}
  \bibinfo{author}{\bibfnamefont{E.~P.~S.} \bibnamefont{Shellard}},
  \bibinfo{journal}{Phys. Rev. Lett.} \textbf{\bibinfo{volume}{92}},
  \bibinfo{pages}{251601} (\bibinfo{year}{2004}), \eprint{hep-ph/0310255}.

\bibitem[{\citenamefont{Battye et~al.}(2009)\citenamefont{Battye, Pearson,
  Pike, and Sutcliffe}}]{BATTYE1}
\bibinfo{author}{\bibfnamefont{R.~A.} \bibnamefont{Battye}},
  \bibinfo{author}{\bibfnamefont{J.~A.} \bibnamefont{Pearson}},
  \bibinfo{author}{\bibfnamefont{S.}~\bibnamefont{Pike}}, \bibnamefont{and}
  \bibinfo{author}{\bibfnamefont{P.~M.} \bibnamefont{Sutcliffe}},
  \bibinfo{journal}{JCAP} \textbf{\bibinfo{volume}{0909}}, \bibinfo{pages}{039}
  (\bibinfo{year}{2009}), \eprint{0908.1865}.

\bibitem[{\citenamefont{Battye and Pearson}(2010)}]{BATTYE2}
\bibinfo{author}{\bibfnamefont{R.~A.} \bibnamefont{Battye}} \bibnamefont{and}
  \bibinfo{author}{\bibfnamefont{J.~A.} \bibnamefont{Pearson}},
  \bibinfo{journal}{Phys.Rev.} \textbf{\bibinfo{volume}{D82}},
  \bibinfo{pages}{125001} (\bibinfo{year}{2010}), \eprint{1010.2328}.

\bibitem[{\citenamefont{Carter}(1997)}]{CARTERB}
\bibinfo{author}{\bibfnamefont{B.}~\bibnamefont{Carter}}
  (\bibinfo{year}{1997}), \eprint{hep-th/9705172}.

\bibitem[{\citenamefont{Avgoustidis and Shellard}(2005)}]{XDIMR}
\bibinfo{author}{\bibfnamefont{A.}~\bibnamefont{Avgoustidis}} \bibnamefont{and}
  \bibinfo{author}{\bibfnamefont{E.~P.~S.} \bibnamefont{Shellard}},
  \bibinfo{journal}{Phys. Rev.} \textbf{\bibinfo{volume}{D71}},
  \bibinfo{pages}{123513} (\bibinfo{year}{2005}), \eprint{hep-ph/0410349}.

\bibitem[{\citenamefont{Kibble and Turok}(1982)}]{KIBTUR}
\bibinfo{author}{\bibfnamefont{T.~W.~B.} \bibnamefont{Kibble}}
  \bibnamefont{and} \bibinfo{author}{\bibfnamefont{N.}~\bibnamefont{Turok}},
  \bibinfo{journal}{Phys.Lett.} \textbf{\bibinfo{volume}{B116}},
  \bibinfo{pages}{141} (\bibinfo{year}{1982}).

\bibitem[{\citenamefont{Burden}(1985)}]{BURDEN}
\bibinfo{author}{\bibfnamefont{C.~J.} \bibnamefont{Burden}},
  \bibinfo{journal}{Phys.Lett.} \textbf{\bibinfo{volume}{B164}},
  \bibinfo{pages}{277} (\bibinfo{year}{1985}).

\bibitem[{\citenamefont{Kibble}(1985)}]{KIB}
\bibinfo{author}{\bibfnamefont{T.~W.~B.} \bibnamefont{Kibble}},
  \bibinfo{journal}{Nucl. Phys.} \textbf{\bibinfo{volume}{B252}},
  \bibinfo{pages}{227} (\bibinfo{year}{1985}).

\bibitem[{\citenamefont{Battye et~al.}(2011)\citenamefont{Battye, Pearson, and
  Moss}}]{BATTYEflaw}
\bibinfo{author}{\bibfnamefont{R.~A.} \bibnamefont{Battye}},
  \bibinfo{author}{\bibfnamefont{J.~A.} \bibnamefont{Pearson}},
  \bibnamefont{and} \bibinfo{author}{\bibfnamefont{A.}~\bibnamefont{Moss}},
  \bibinfo{journal}{Phys.Rev.} \textbf{\bibinfo{volume}{D84}},
  \bibinfo{pages}{125032} (\bibinfo{year}{2011}), \eprint{1107.1325}.

\bibitem[{\citenamefont{Martins and Shellard}(2006)}]{Fractal}
\bibinfo{author}{\bibfnamefont{C.~J. A.~P.} \bibnamefont{Martins}}
  \bibnamefont{and} \bibinfo{author}{\bibfnamefont{E.~P.~S.}
  \bibnamefont{Shellard}}, \bibinfo{journal}{Phys.Rev.}
  \textbf{\bibinfo{volume}{D73}}, \bibinfo{pages}{043515}
  (\bibinfo{year}{2006}), \eprint{astro-ph/0511792}.

\end{thebibliography}
\end{document}